\documentclass[12pt]{article}
\usepackage{graphicx}
\usepackage{setspace}
\usepackage{epstopdf}
\usepackage{bm}
\usepackage{float}
\usepackage{amssymb}
\usepackage{amsmath}
\usepackage{cite}

\newcommand{\spdn}{|\mkern -4mu \downarrow\mkern 0.5mu \rangle}
\newcommand{\spup}{|\mkern -4mu \uparrow \rangle}

\begin{document}
\singlespace
\title{An approach to quantum gas microscopy of polar molecules}
\author{Jacob P. Covey*, Luigi De Marco, \'{O}scar L. Acevedo,\\
\\Ana Maria Rey, and Jun Ye\\
\\
\normalsize{JILA, National Institute of Standards and Technology and University of Colorado,} \\ \normalsize{and Department of Physics, University of Colorado, Boulder, CO 80309, USA}
\\
\normalsize{*Corresponding author: jacob.covey@colorado.edu}}
\date{}

\maketitle

\begin{abstract}
\singlespace
Ultracold polar molecules are an ideal platform for studying many-body physics with long-range dipolar interactions. Experiments in this field have progressed enormously, and several groups are pursuing advanced apparatus for manipulation of molecules with electric fields as well as single-atom-resolved \textit{in situ} detection. Such detection has become ubiquitous for atoms in optical lattices and tweezer arrays, but has yet to be demonstrated for ultracold polar molecules. Here we present a proposal for the implementation of quantum gas microscopy for polar molecules, and specifically discuss a technique for spin-resolved molecular detection. We use numerical simulation of spin dynamics of lattice-confined polar molecules to show how such a scheme would be of utility in a spin-diffusion experiment.
\end{abstract}

\clearpage\doublespace
\section{Introduction}
Ultracold polar molecules present an idyllic platform for emulating quantum magnetism in many-body long-range interacting systems~\cite{Micheli2006,Carr2009,Kwasigroch2017,Gorshkov2011b}. This requires precise state preparation~\cite{Yan2013,Moses2015}, large DC electric fields to control the strength of the Ising interaction, and precise read-out based on rotational spectroscopy and/or high resolution \textit{in situ} detection to measure spatial correlations~\cite{Weitenberg2011}. Such high resolution detection of individual particles has become a standard tool in cold atom experiments.

Quantum gas microscopy of atoms is made possible through high-resolution imaging in a deep optical lattice that freezes atomic motion during imaging, along with laser cooling which prevents atoms from heating while scattering many photons~\cite{Bakr2009,Sherson2010}. However, such an imaging process scrambles the hyperfine state of the atoms, thereby rendering their spin degree of freedom undetectable. This issue has been circumvented using various techniques ~\cite{Preiss2015b,Boll2016, Kwon2017}, but typically one atomic spin state is removed with a short pulse of resonant light prior to detecting the other spin state~\cite{Cheuk2016a,Weitenberg2011,Greif2016}. 

In ultracold molecule quantum simulators, the molecular rotational state is used as the analogue of magnetic spin~\cite{Gorshkov2011b,Yan2013}. Efforts towards \textit{in situ} single molecule detection in both optical lattices~\cite{Gempel2016,Grobner2016,Covey2017t} and optical tweezers~\cite{Hutzler2016,Liu2017} are increasingly active. It is thus timely to consider quantum gas microscopy of polar molecules, and how two rotational states can be simultaneously detected. The latter point is of particular importance for molecules due to the limited fidelity of their creation process, which results in the occupation of a given site being \textit{a priori} unknown~\cite{Moses2015,Covey2016}. The ability to detect the presence or absence of a molecule on a given site, in addition to its state, is therefore important.

\begin{figure}[ht]
\centering
\includegraphics[width=16cm]{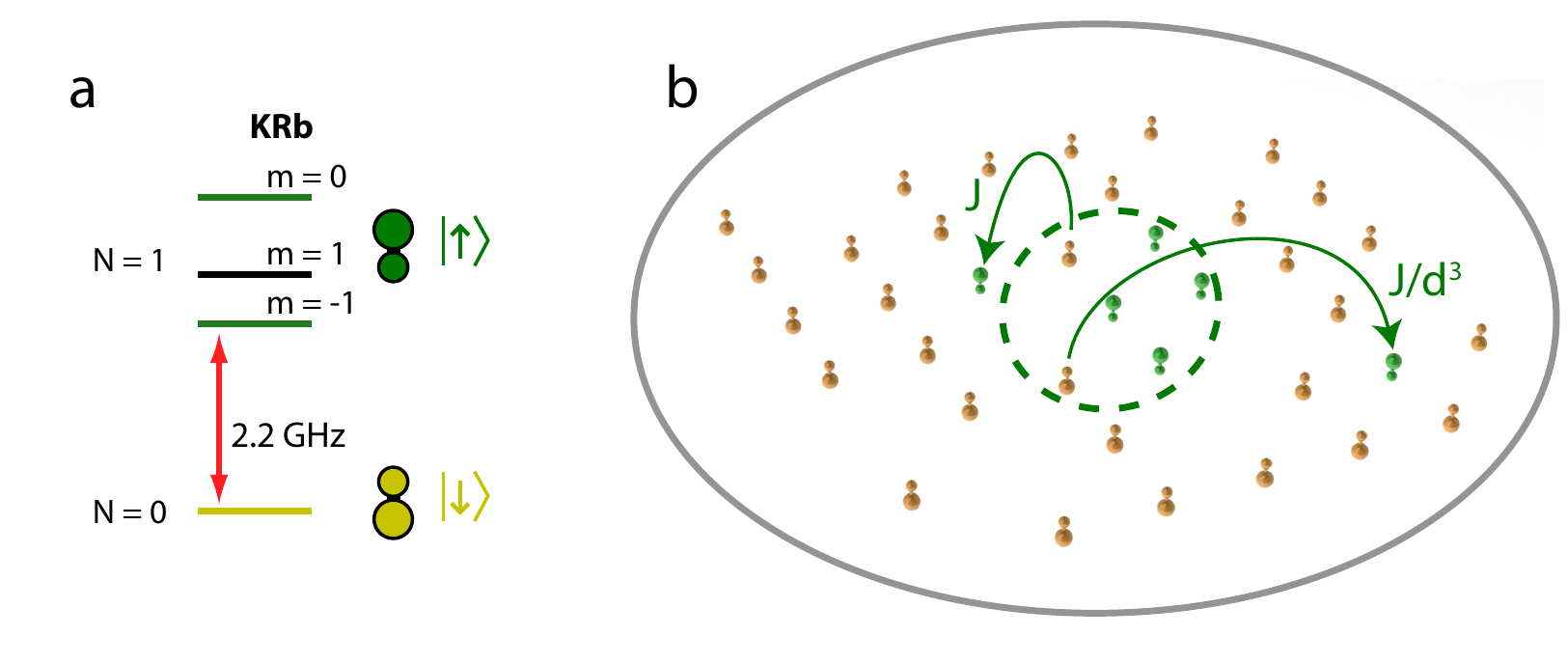}
\caption{A quantum gas microscope for polar molecules. (a) The rotational energy levels of KRb. (b) Illustration of spin propagation on a 2D plane of molecules following a local excitation within the green dashed region.}
\label{fig:Figure1}
\end{figure}

In this paper, we present a technique by which two molecular rotational states can be unambiguously detected, giving simultaneously site-resolved and spin-resolved detection. After discussing the approach in which molecules are detected, we describe how these techniques can be extended to fluorescence detection for quantum gas microscopy. Finally, we present simulations of out-of-equilibrium spin systems, and describe how the investigation of such dynamics would proceed under a spin-resolved quantum gas microscope. $^{40}$K$^{87}$Rb~\cite{Ni2008} is used as an example throughout, but these methods are general, and they can be applied to other ultracold bialkali molecular species~\cite{Park2015,Seesselberg2017,Takekoshi2014,Molony2014,Guo2016}.

\section{Spin-resolved molecular microscopy}
Ultracold polar molecules are produced by optically transferring weakly-bound, magneto-associated Feshbach molecules~\cite{Zirbel2008,Chin2010} to the absolute rovibronic ground state using STImulated Raman Adiabatic Passage (STIRAP)~\cite{Ni2008,Danzl2010}. Imaging molecules directly is challenging since there are no closed cycling transitions as in atoms, though direct imaging has been accomplished with a limited signal-to-noise ratio~\cite{Wang2010}. However, by reversing the STIRAP and magneto-association processes, a ground-state polar molecule can be converted to a pair of free atoms with $\sim$90--95\% fidelity~\cite{Ni2008,Zirbel2008,Covey2016,Aikawa2009,Vitanov2016,Moses2017,Covey2017a}, which can then be imaged on cycling transitions. 

The first few rotational states for KRb are shown in Fig.~\ref{fig:Figure1}a. The states are denoted by $|N,m_N \rangle$, where $N$ is the rotational quantum number, and $m_N$ is its projection onto the quantization axis (i.e. electric or magnetic field). As shown in Fig.~\ref{fig:Figure1}b, site-resolved microscopy can be used to perform spin-diffusion studies in a 2D plane. After an initial preparation of spin excitation within a local region, the excitation can diffuse throughout the system via dipolar interactions characterized by an energy scale $\hbar\textit{J}$, where $\hbar$ is Planck's constant divided by $2\pi$, that falls off in range as $r^{-3}$.

\subsection{Creating a two-dimensional molecular sample}
For quantum gas microscopy, it is important to prepare the system in a two-dimensional geometry for high-resolution, single-site imaging. With ultracold atoms this can be accomplished by either using a magnetic field gradient to spectroscopically select a single layer with a hyperfine transition~\cite{Sherson2010,Cheuk2015}, or by using an accordion lattice~\cite{Mitra2016}. Similar options are available for ultracold molecule experiments. Our favored approach is to spectroscopically select a single layer of molecules using a rotational transition. This can be done by applying an electric field gradient which generates a different DC Stark shift for each layer.

\subsection{Spin-resolved imaging protocol}
Formation and dissociation of ground-state molecules in an optical lattice is well understood~\cite{Moses2015,Covey2016,Chotia2012}, and it has been demonstrated that either K or Rb atoms from dissociated KRb molecules can be imaged \textit{in situ} to yield consistent results~\cite{Moses2015,Covey2016}. 
This is an important step towards building a spin-resolved quantum gas microscope of polar molecules as this requires both atoms to be a suitable proxy for molecules. It has also been shown, equally as important, that one species can be removed with resonant light without deleterious effects on the other. In fact, removing a unique atomic species is straightforward since the difference in transition frequencies is on the order of several THz.

Accordingly, each spin state can be mapped onto either atomic species, making a spin-resolved quantum gas microscope of KRb molecules equivalent to simultaneous quantum gas microscopes of K~\cite{Haller2015,Cheuk2015,Edge2015} and Rb~\cite{Bakr2009,Sherson2010}. We also note that KRb has the convenient feature that the two $\text{S}_{1/2}\rightarrow \text{P}_{3/2}$ atomic transitions are separated by only 13~nm, which minimizes chromatic aberrations in imaging while still making wavelength separation straightforward.

\begin{figure}[ht]
\centering
\includegraphics[width=13cm]{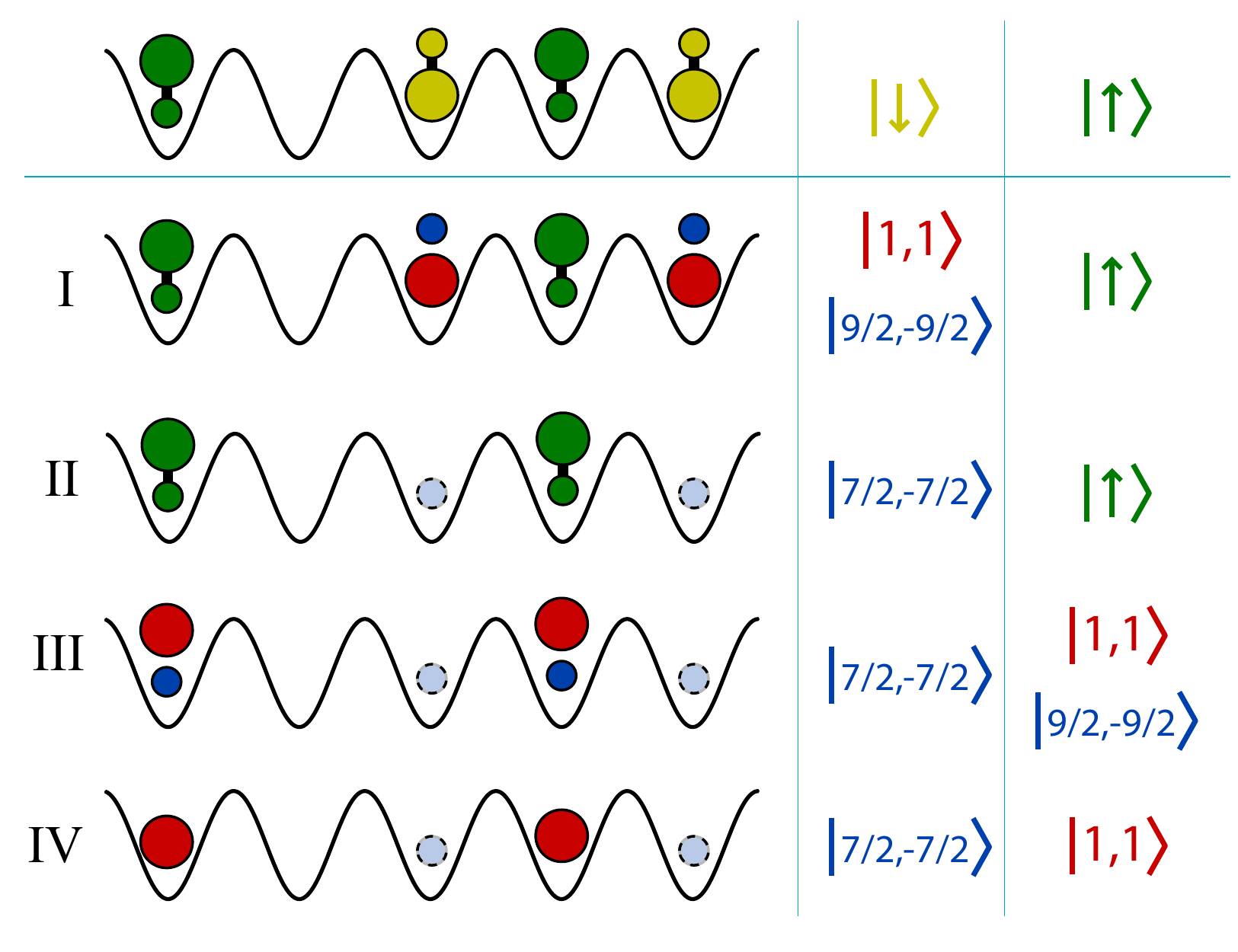}
\caption{Various steps required to realize a spin-resolved microscope, mapping $\spdn$ onto K and $\spup$ onto Rb. The left column shows the color-coded particles on lattice sites, and the right columns describe the colors as specific states of KRb molecules and K and Rb atoms. The top row shows the initial configuration with $\spdn$ and $\spup$ molecules in the lattice. The subsequent rows describe the steps which result in the bottom row, where only K or Rb atoms remain on lattice sites that were initially populated with a $\spdn$ or $\spup$ molecule, respectively, as in the top row. See the text for an explanation of each step.}
\label{fig:Figure2}
\end{figure}

Before describing the mapping protocol in detail, we note that there is freedom in which molecular state is mapped onto which atomic species. Either option is possible in general, but for a given molecular species one choice may prove to be advantageous over the other. In what follows, we focus on the case in which $\spdn$ is mapped onto K, as shown in Fig.~\ref{fig:Figure2}. We also note that we assume the lattice is sufficiently deep such that the tunneling probabilities for molecules or atoms are negligible throughout the entirety of this procedure.

\subsubsection*{Step I: Dissociating $\spdn$ molecules}

The first step in the mapping protocol is to convert $\spdn$ molecules to a pair of K and Rb on the same site. This is done by reversing the STIRAP sequence, which uniquely couples the $\spdn$ state to Feshbach molecules, and then sweeping the magnetic field across the Feshbach resonance to dissociate the Feshbach molecules. The fidelity of this process is $\sim90-95\%$~\cite{Covey2016,Ni2008}, which is limited by the STIRAP efficiency. After dissociation, Rb is in the $|F,m_F\rangle=|1,1\rangle$ state and K is in the $|9/2,-9/2\rangle$ state, as shown in step I of Fig.~\ref{fig:Figure2} ($F$ denotes the total angular momentum, and $m_F$ is its projection on the magnetic field axis).

This STIRAP pulse will not couple $\spup$ molecules to the specific Feshbach state, but we must ensure that it does not couple them to any other state. Accordingly, we must look carefully at the full quantum state of the molecules, including the nuclear hyperfine degrees of freedom~\cite{Ospelkaus2010a}. In a STIRAP sequence with identically polarized beams, the molecule's angular momentum is conserved, which eliminate most pathways. Furthermore, the STIRAP linewidth is $\sim$200~kHz and the sequence is 5~$\mu$s long~\cite{Ni2008}, and since the rotational splitting is on the order of GHz, the spectral resolution prohibits driving to an uncoupled state. An electric field can additionally be used during STIRAP to provide even greater selectivity~\cite{Ni2008}.

\subsubsection*{Step II: Removing Rb atoms}
Following dissociation, one of the species must be removed such that each doublon becomes a site with a single atom. As discussed above, this can be done with a pulse of resonant light. Such pulses, which have been demonstrated to preserve the other atomic spin with high fidelity, are routinely used in atomic quantum gas microscope experiments, with typical blast times between 10 $\mu$s and 1 ms depending on the species and lattice depth~\cite{Moses2015,Parsons2016,Brown2016,Cheuk2016b,Bakr2009,Sherson2010,Haller2015,Cheuk2015,Edge2015}. Moreover, pulses resonant with the atomic transition have a negligible effect on molecules~\cite{Covey2016,Ni2008,Moses2015}, thus leaving the $\spup$ molecules unaffected.

The choice of which atom to be removed first depends on the particular molecular species. Here we consider the removal of Rb first, as in Fig.~\ref{fig:Figure2}. Effective removal of Rb can be accomplished in $\sim$1~ms with $>99.9\%$ fidelity by simultaneously driving the $|1,1\rangle\rightarrow|2,2\rangle^\prime$ and $|2,2\rangle\rightarrow|3,3\rangle^\prime$ transitions, where $|F,m_F\rangle^\prime$ denotes the electronically excited $P_{3/2}$ state. This leaves behind K atoms in the $|9/2,-9/2\rangle$ state.

With Rb atoms removed, we must flip the K atoms to a different spin state such that they can be distinguished from the K atoms that emerge in the subsequent steps. For K, a convenient transition is from $|9/2,-9/2\rangle$ to $|7/2,-7/2\rangle$, which has a transition frequency of $\sim$2.7~GHz under 550 G; this transition can be driven with $>$$99\%$ fidelity in tens of $\mu$s~\cite{Cheuk2016b}. As shown in Fig.~\ref{fig:Figure2}, after step~II, molecules in the $\spup$ state and unpaired K atoms remain, with the latter depicted in lighter color with hashed edges to signify the $|7/2,-7/2\rangle$ state.

\subsubsection*{Step III: Dissociating $\spup$ molecules}
We next convert the $\spup$ molecules to a doublon, first by applying a rotational-state-changing microwave pulse to transfer $\spup\to\spdn$. Such pulses have $>99\%$ fidelity in tens of $\mu$s~\cite{Yan2013}. Then the same reverse-STIRAP and magneto-dissociation sequence converts these molecules to doublons. As before, the doublons are in the states $|1,1\rangle$ for Rb and $|9/2,-9/2\rangle$ for K. This is shown in step~III of Fig.~\ref{fig:Figure2}.

\subsubsection*{Step IV: Removing K atoms}
In the final step, K atoms from the doublons are removed, which shows why the atomic state transfer was necessary in step~II. The K atoms that remain from the previous step are in the $|7/2,-7/2\rangle$ state, which are sufficiently detuned from the $|9/2,-9/2\rangle\to|11/2,-11/2\rangle^\prime$ transition as to not be affected. Therefore, the K atoms in the doublon can be removed with a blast pulse in $<1$ ms with high fidelity~\cite{Cheuk2016b}, which completes the mapping procedure.

\subsection{Single-molecule addressing}
\begin{figure}[ht]
\begin{center}
\includegraphics[width=16cm]{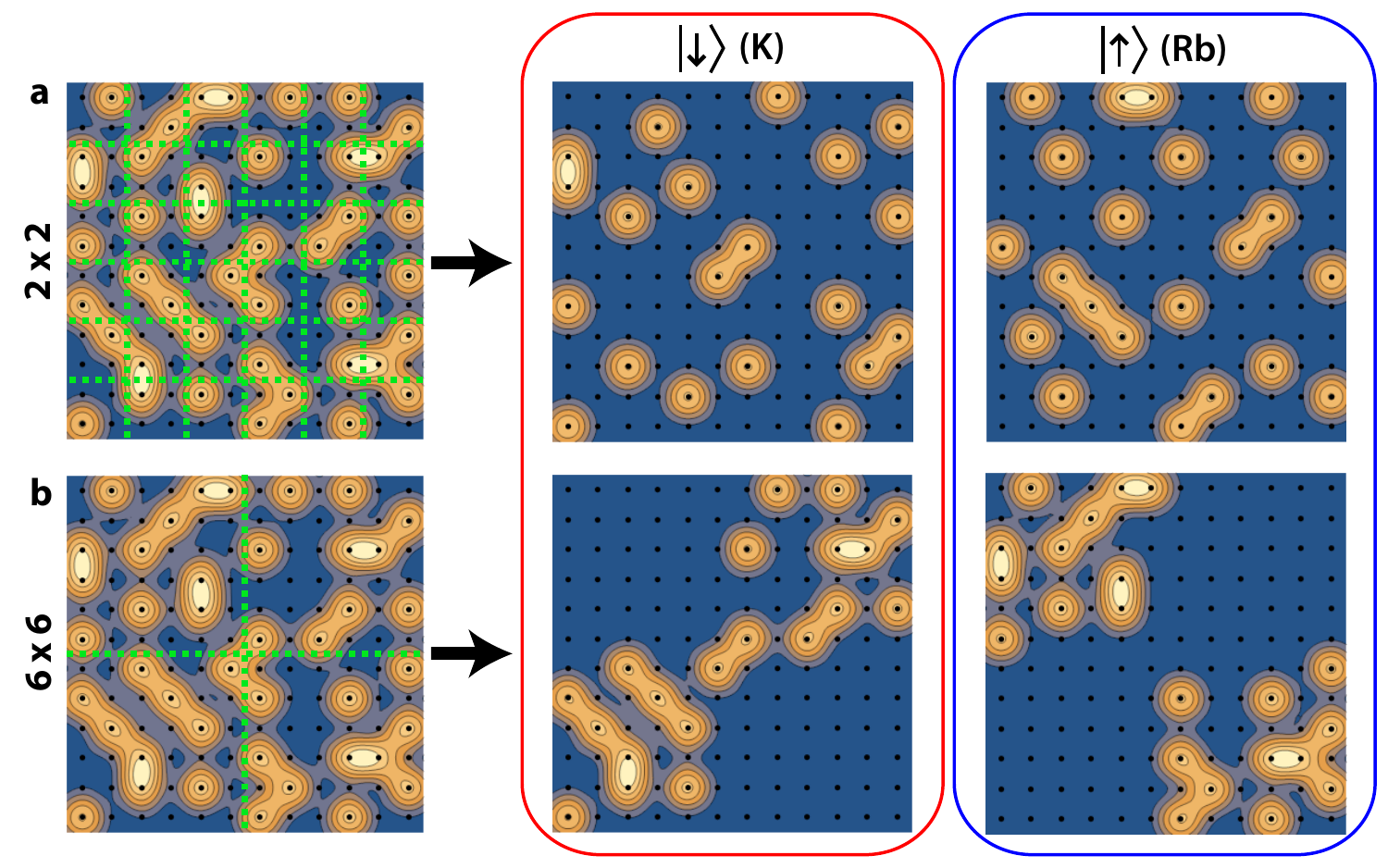}
\caption{Illustration of spin-resolved microscopy. A  lattice of molecules with $\approx30\%$ filling is imaged with a resolution of $R_\text{min}=800$ nm. Black dots denote the lattice sites, and the yellow intensity distributions are the point spread functions of the atoms with this resolution. The light blue lines on the left show the checkerboard pattern that is used to create the desired spin configuration, which corresponds to $2\times2$ checkers in (a) and $6\times6$ checkers in (b). The right images show $\spdn$ only and $\spup$ only for both checker sizes.}
\label{fig:Figure3}
\end{center}
\end{figure}

The new KRb apparatus at JILA~\cite{Covey2017t} has a microscope objective for high-resolution detection of polar molecules, in addition to in-vacuum electrodes for single layer selection. Currently, an objective with numerical aperture (NA) of 0.53 is being used, which corresponds to an optical resolution of $R_\text{min}=900$~nm at $\lambda=780$~nm. The lattice spacing is 532 nm. However, future work may use a NA as high as 0.65, with a resolution of 730 nm. A lattice with a $f=30\%$ filling fraction is shown in Fig.~\ref{fig:Figure3}, and the sites with molecules are represented as the point spread function corresponding to a representative resolution of $R_\text{min}=800$~nm.

A beam sent through the objective can generate a tight focus on the molecules, which can impart a large AC Stark shift to address a particular molecule. At $\lambda=1064$~nm the molecules are not resonantly coupled from the rovibronic ground state to the electronic excited potentials, resulting in a low scattering rate~\cite{Chotia2012} while still providing a large polarizability~\cite{Neyenhuis2012}. To uniquely select the molecules within the addressing beam, we require that the AC Stark shift be large compared to the Rabi frequency used to drive the rotational transition, which is typically $\sim$10~kHz. For a beam of $1/e^2$ radius $\approx1$ $\mu$m, a power of only $\approx10$ $\mu$W is needed for a 50 kHz differential AC Stark shift of the $|N,m_N\rangle=|0,0\rangle-|1,0\rangle$ transition at the center of the beam. 

The prototypical case to consider is a round $\spup$ region as shown in Fig.~\ref{fig:Figure1}b. However, our simulations suggest that the perimeter of the $\spup$ region must be a significant fraction of the enclosed sites in order to see a decay of spin diffusion within experimentally relevant timescales of a few $1/\textit{J}$ (see Fig.~\ref{fig:Figure4}). Therefore, a single circular $\spup$ region of 10's of sites (as in Fig.~\ref{fig:Figure1}b) would have a small boundary perimeter compared to the total number of sites, and hence spin diffusion would be quite limited and slow. Instead it is desirable to create a scenario where the spin domain perimeter is a significant fraction of the total spins.

To go beyond a single tightly-focused beam, we will introduce an arbitrary potential imaged onto the molecules by way of a spatial light modulator, such as a digital micro-mirror device (DMD) (see Fig.~\ref{fig:Figure3}). These techniques have been used to create arbitrary potentials in atomic systems~\cite{Choi2016,Mazurenko2016,Preiss2015}, and could be applied to molecules to create a disordered spin distribution in the lattice as a platform for investigating, for example, many-body localization~\cite{Yao2014}. 

As discussed in Section 3, we are particularly interested in the case of a checkerboard spin pattern as our initial condition for spin diffusion experiments. The two cases in Fig.~\ref{fig:Figure3} correspond to checker sizes of $2\times2$ and $6\times6$ sites. Note that when the checker is sufficiently small as in Fig.~\ref{fig:Figure3}a, there is $<1$ molecule of a given spin in each checker, and it is thus difficult to track the dynamics when detecting only one spin. This shows that spin-resolved detection is essential in such scenarios.

\subsection{Detection fidelity and technical requirements}
After the above protocol, the sites that were $\spdn$ molecules are K atoms in $|7/2,-7/2\rangle$ and the sites that were $\spup$ molecules are Rb atoms in $|1,1\rangle$. The subsequent imaging via optical molasses~\cite{Bakr2009,Sherson2010} cooling or Raman sideband cooling~\cite{Patil2014} (RSC) operates primarily on the transitions $|2,2\rangle\to|3,3\rangle^\prime$ for Rb and $|9/2,-9/2\rangle\to|11/2,-11/2\rangle^\prime$ for K, but the repump transitions $|1,1\rangle\to|2,2\rangle^\prime$ for Rb and $|7/2,-7/2\rangle\to|9/2,-9/2\rangle^\prime$ for K are required anyway to keep atoms in the cycling transition~\cite{Bakr2009,Sherson2010,Haller2015,Cheuk2015,Edge2015}. Hence, both transitions are driven in parallel, and the initial states of the K and Rb atoms in the lattice are unimportant for imaging.

The particular choice of a mapping protocol hinges on the fact that two atoms on a lattice site will quickly undergo inelastic loss when one occupies an excited hyperfine level~\cite{Covey2016}. Therefore, the time spent in this configuration must be minimized. Due to the inverted hyperfine structure of K, the ground hyperfine level is the cycling state, and thus it is convenient to remove K atoms in Step IV as opposed to Rb. In cases where neither atom has an inverted hyperfine structure, such as NaRb or RbCs, the protocol with higher overall fidelity would involve removing the lighter atom in Step IV.

It is critical that the desired atomic species is preserved during removal of the other. One possible deleterious effect is heating of the desired species by the removal pulses. The heating rates are calculated using the photon scattering rates given by:
\begin{equation}
\Gamma_\text{sc}=\left(\frac{\Gamma}{2}\right)\frac{(I/I_\text{sat})}{1+4(\Delta/\Gamma)^2+(I/I_\text{sat})},
\end{equation}
where $\Gamma$ is the linewidth of the electronic excited state, $I$ is the intensity of the removal beam, $I_\text{sat}$ is the saturation intensity of the atomic transition, and $\Delta$ is the detuning from resonance of the excited state. The heating rate can be calculated from this scattering rate via $\dot{T}=1/3 \times (2 \times E_\text{rec})/k_\text{B} \times \Gamma_\text{sc}$, where the $1/3$ is specific to the case of a harmonic trap, and $2\times E_\mathrm{rec}$ is the recoil energy of photon absorption and re-emission~\cite{Grimm2000}.

For a removal beam of intensity $I= I_\text{sat}$ and detuning $\Delta=0$ for the target species, the scattering rate is $\Gamma_\text{sc} \approx 10^7$~$\text{s}^{-1}$, which corresponds to a resonant heating rate of $\approx$1~$\mu$K/$\mu$s for Rb in the $|1,1\rangle$ state and $\approx$2~$\mu$K/$\mu$s for K in $|9/2,-9/2\rangle$.  These on-resonance heating rates are strong such that the removal pulse can be quite short, on the scale of 1 ms or less, and consequently do not perturb the desired species or spin states for which the off-resonance heating rates are very low. During Step II, for example, a resonant pulse to remove Rb atoms will have no impact on either K atoms or KRb molecules. In Step IV, we need to remove K atoms in the $|9/2,-9/2\rangle$ state without affecting the $|7/2,-7/2\rangle$ state. For the removal light tuned on resonance with $|9/2,-9/2\rangle\to|11/2,-11/2\rangle^\prime$, the $|7/2,-7/2\rangle\to|9/2,-9/2\rangle^\prime$ transition has the smallest detuning of 0.84~GHz, corresponding to a scattering rate of $\Gamma_{|7/2,-7/2\rangle\to|9/2,-9/2\rangle^\prime} = 245$~$\mathrm{s}^{-1}$. Meanwhile, a pulse of length $\sim1$~ms is more than sufficient to remove $|9/2,-9/2\rangle$ K atoms.

Another factor that may play a role during the removal pulses is photo-association. Quantum gas microscopes operate with parity detection because every pair of atoms located on a single lattice site is associated into an electronically excited molecule during the roughly second-long interrogation time. However, the probability of this process happening during the $\sim$ms removal pulses is known to be low. 

We can now estimate the overall detection fidelity of each individual spin component. The detection of $\spdn$ relies on STIRAP, Rb removal, and a K spin flip, which have respective efficiencies of $95\%$, $>99\%$, and $>99\%$. This corresponds to an overall fidelity of $95\%$. The detection of $\spup$ relies on STIRAP and K removal (while preserving K atoms of the other hyperfine state), which have respective efficiencies of $95\%$ and $>95\%$~\cite{Cheuk2016b}. However, it also relies on minimal deleterious effects from the first STIRAP pulse. Such effects are expected to be at the single percent level, but will vary between molecular species. Therefore we estimate the overall fidelity of $\spup$ detection to be $85-90\%$.

\section{Spin impurity dynamics}
To further motivate the need for a checkerboard spin pattern and for spin-resolved microscopy, we present simulations of spin-exchange dynamics mediated by long-range dipolar interactions, which give rise to interesting and complex spatial correlations. The Hamiltonian that describes this exchange process is~\cite{Gorshkov2011b,Wall2015,Yan2013}:
\begin{equation}
\hat{H} = \frac{\hbar J}{2}\sum_{j=1}^\mathcal{N}\sum_{l<j}(1-3\text{cos}^2\theta_{lj})||\boldsymbol{r}_l-\boldsymbol{r}_j||^{-3}\left(\hat{s}_l^+\hat{s}_j^-+\hat{s}_l^-\hat{s}_j^+\right),
\end{equation}
where $\hat{s}_j$ are spin--1/2 operators, $\boldsymbol{r}_j$ is the position vector of molecule $i$ in the lattice, $\mathcal{N}$ is the number of molecules, and $\theta_{lj}$ is the angle between the vector $\boldsymbol{r}_l-\boldsymbol{r}_j$ and the quantization axis, which is defined by an external magnetic field. This Hamiltonian is derived for the case of zero electric field at DC so that there is no Ising term proportional to $\hat{s}^z_l\hat{s}^z_j$. The exchange coupling $J$ is determined by the transition dipole moment, $d_{\uparrow\downarrow}$, between two selected rotational states and is given by $\hbar J=d^2_{\uparrow\downarrow}/(4\pi\epsilon_0a^3)$, where $\epsilon_0$ is the permittivity of free space and $a$ is the lattice spacing.

For KRb molecules in a three-dimensional lattice with spacing $a=532$ nm, using the two rotational states $|0,0\rangle$ and $|1,0\rangle$, $d_{\uparrow\downarrow}\approx-0.57/\sqrt{3}$~D, and $|J|/2=2\pi\times104$~Hz. For $|0,0\rangle$ and $|1,-1\rangle$, $|J|/2=2\pi\times52$~Hz. We assume that the quantization axis is perpendicular to the two-dimensional plane of molecules, which implies that $\text{cos}~\theta_{lk}=0$, resulting in an isotropic interaction. Disorder in such a model manifests itself in two ways: either a disordered potential landscape can be added to the lattice using a projected potential as discussed in the previous section; or, since the lattice filling fraction is less than unity, there is a natural disorder in the filling arrangement of the molecules and consequently in the $||\boldsymbol{r}_l-\boldsymbol{r}_j||^{-3}$ geometrical prefactors for dipolar coupling. The question about the existence of many-body localization in this 2D system remains open and largely unexplored~\cite{Peter2012,Kwasigroch2014,Yao2014,Burin2015,Burin2015a,Kwasigroch2017}, while the single particle (Anderson localization) case is better understood~\cite{Xu2015,Deng2016}.

The initial state is prepared by driving all molecules to the $\spdn$ state and then selecting an excitation region of the lattice in which the molecules are excited to $\spup$. Once prepared, the total spin $\hat{S}_z$ remains constant since it is a conserved quantity. In this case, a useful observable for quantifying the spin dynamics is the \textit{imbalance} $\hat I_z(t)$, which is a measure of the average magnetization of the gas weighted by the initial onsite values, defined as
\begin{equation}
\hat{I}_z(t)=\frac{4}{\mathcal{N}}\sum_{j=1}^\mathcal{N}\langle\hat{s}_j^z(0)\rangle\hat{s}_j^z(t).
\end{equation}
The imbalance is equal to one initially and will either decay if the system is ergodic or will remain finite at long times if the system localizes.

\begin{figure}[ht]
\begin{center}
\includegraphics[width=16cm]{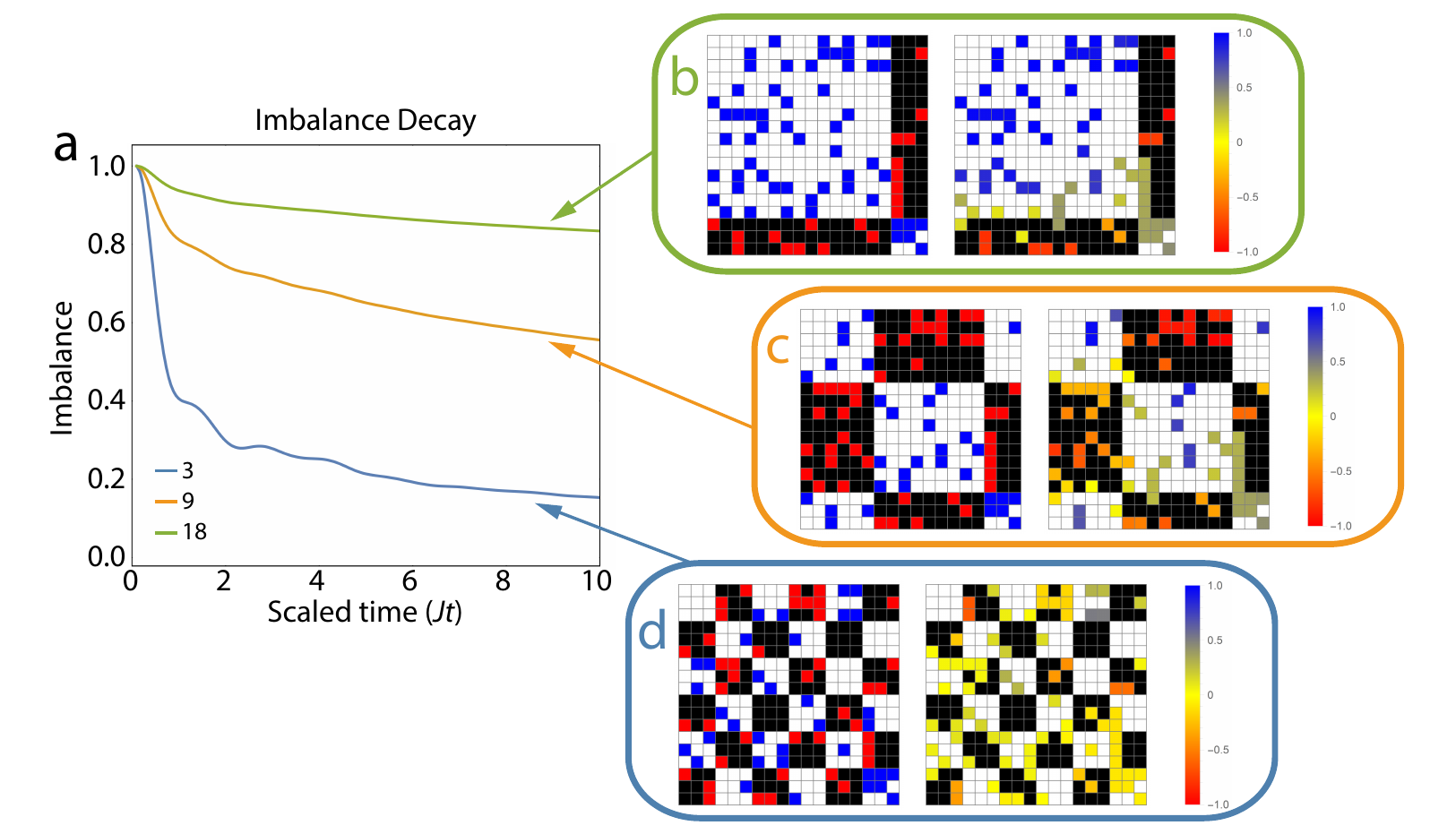}
\caption{Numerical DTWA simulations of spin dynamics. (a) The spin imbalance as a function of time for $36\times36$ sites for checker sizes of $3\times3$ (blue), $9\times9$ (orange), and $18\times18$ (green) sites. (b)-(d) Spin-resolved microscopy of an $18\times18$ subset with 25\% filling at $t=0$ (left) and $Jt=10$ (right) corresponding to the three curves in (a). Sites on checkers that are initially $\spup$ ($\spdn$) are colored white (black). The color scale is from blue to red for $\spup$ to $\spdn$, respectively.}
\label{fig:Figure4}
\end{center}
\end{figure}

Since the exact simulation of interacting spins in a 2D lattice becomes intractable as the number of molecules is increased, the simulations presented here are based on the discrete truncated Wigner approximation (DTWA)~\cite{Schachenmayer2015,Schachenmayer2015a,Pucci2016}. The DTWA is a semi-classical method that models the dynamics through a set of classical trajectories (in this case Bloch vector dynamics) that evolve according to the mean field equations. The random initial conditions for each of these trajectories is selected according to the initial state represented as a quasi-probability Wigner distribution. Despite its semiclassical character, DTWA has been shown to be capable of reproducing quantum correlations and to capture the quantum spin dynamics beyond the mean field limit; it has been demonstrated recently to be useful for exploring ergodic/localized dynamics~\cite{Acevedo2017}.

We now consider the case where the initial excitation is a checkerboard pattern (which can be generated with spatial light modulators as discussed in the previous section), where the size of each square can be as small as $2 \times 2$ sites (see Fig.~\ref{fig:Figure3}). For large checker sizes the spin imbalance stays near unity even out to long times, in accordance with the case of a single large circular $\spup$ region. However, as the checker size decreases, substantial spin dynamics occurs and the imbalance drops to $\approx0.2\%$ within $t\approx2/J$, as in Fig.~\ref{fig:Figure4}a. 

Snapshots of the spin distributions at $t=0$ and $t=10/J$ for checker sizes of $3\times3$, $9\times9$, and $18\times18$ are shown for Fig.~\ref{fig:Figure4}b-d. It is also worth noting that under our initial conditions total spin observables $\hat{S}_\alpha$ ($\alpha=x,y,z$) have expectation values equal to zero throughout the entire dynamic process. Therefore, site- and spin-resolved detection is essential to glean all the information from such dynamics. These numerical predictions provide a strong motivation for the experimental capability of microscopic, spin-resolved detection since substantial diffusion occurs only for microscopic areas.

\section{Outlook}
An investigation complementary to such spin diffusion experiments is to spectroscopically track the dynamics after driving the molecules to the equatorial plane of the spin Bloch sphere ($|\mkern -4mu \to\rangle=1/\sqrt{2}(\spdn+\spup)$. This provides a means of studying the so-called Bose-Einstein condensed (BEC) phase of rotational excitations, which has been predicted to appear in 2D above a critical filling fraction~\cite{Peter2012,Kwasigroch2014,Kwasigroch2017}. The BEC phase manifests as a divergent $T_2$ coherence time of the rotational transition even in the presence of inhomogeneous broadening. 

The underlying physics behind the long coherence time is the relativistic dispersion relation of the spin wave excitations in 2D for small values of the wave vector, $\delta\epsilon_\textbf{k}\propto|\textbf{k}|$, which has striking consequences for the thermodynamic phase diagram~\cite{Peter2012,Kwasigroch2014,Kwasigroch2017}. We expect that topological excitations such as vortices and solitonic rings could be imprinted and observed in such a system with the aid of microscopic addressing and detection. Furthermore, entanglement of superfluid twins may allow for microscopic measurements of the entanglement entropy in a long-range-interacting system~\cite{Islam2015}.

\subsection*{Acknowledgments}
We would like to acknowledge fruitful discussions with Steven Moses, Giacomo Valtolina, Kyle Matsuda, William Tobias, Robert Lewis-Swan, Arghavan Safavi-Naini, and Rahul Nandkishore. We acknowledge critical reading of the manuscript by Collin Kennedy, Edward Marti, and Manuel Endres. We acknowledge funding from NIST, AFOSR-MURI Advanced Quantum Materials, NSF-PHYS 1734006, NSF-PHY 1521080, AFOSR FA9550-13-1-0086, and ARO-MURI Cold Molecules.

\bibliographystyle{unsrt}

\end{document}